\documentstyle[12pt]{article}

\begin{document}

\author{Guang-jiong Ni \thanks{Email: gjni@fudan.ac.cn} \\
Department of Physics, Fudan University \\ Shanghai
200433, P. R. China}
\title{To Enjoy the Morning Flower in the Evening------Where is the Subtlety of
Quantum Mechanics?
\thanks{%
This is an English version of its original article published in the Chinese
journal {\it Kexue (Science)} Vol. 50, No. 2, 38$\sim $42,
1998}}

\date{The Date }
\maketitle

\begin{abstract}
Why does the $i=\sqrt{-1}$ appear essentially in the quantum mechanics?
Why are there operators and noncommutativity (the uncertainty relation)
in the quantum mechanics? Why are these two aspects closely related and
indivisible? In probing these problems, a new point of view is proposed
tentatively.
\end{abstract}

Dirac, who had made prominent contribution to the establishment and
development of Quantum Mechanics ( QM ), said in 1970 that in the past he
deemed the ``noncommutativity'' characterized by the Planck constant $%
\hbar$ being the main feature of QM, but after 1970 he tended to consider
the existence of probability amplitude, i.e., the wave function as the most
important one. He further pointed out that where the appearance of phase in
wave function looks the most subtle.

In C.N. Yang's thinking, both two above features are very important and form
the revolutionary progress with far-reaching meaning in the description of
nature. Moreover, he pointed out that these two aspects are related
intimately and there is an imaginary unit $i=\sqrt{-1}$ appearing in them
[1].

\section{Why $i=\protect\sqrt{-1}$ appears in QM?}

Being one of the founders of QM, Schr\"{o}dinger proposed his famous
equation by writing six papers successively. Among them all the weal and woe
are closely around an important problem whether it is necessary to introduce
the imaginary unit $i=\sqrt{-1}$ ?[1] Eventually, when he made up his mind
to write it in, he brought an epoch-making reform into physics.

It is well known that in classical physics or electric engineering the use
of $i$ is rather convenient. If the alternative current is denoted by $%
I(t)=I_0e^{i\omega t}$ , it is quite easy to do calculation in calculus.
Finally one has to pick the real part of result corresponding to the
measured quantity.

However, the introduction of $i$ in QM is by no means a matter of
convenience but a need in essence. If discarding $i$ , the Schr\"{o}dinger
equation would reduce into a classical equation similar to that describing
the heat conduction or particle diffusion, it would be totally impossible to
describe the so-called ``wave particle dualism'' of microscopic particles.

Today we already understand that the most important or subtle aspect in QM
is the phase of wave function, which in turn must be expressed by $i$ .
Though the Born statistical interpretation was in conformity with the
experiments, the phase was erased and the $i$ disappeared too. Despite the
brilliant success of QM, its fundamental interpretation remains still in a
rather obscure status. This is a seldom phenomenon in the history of natural
science. How long time need we further to wait for? Let me try to propose
some personal understanding for the reference of our readers.

\section{The phase transformation of wave function}

One used to discuss the phase transformation of wave function $\psi $ in QM
as follows:

\begin{equation}
\psi \longrightarrow \psi ^{^{\prime }}=e^{i\theta }\psi
\end{equation}

\[
{\rm Re}\psi \longrightarrow {\rm Re}\psi ^{^{\prime }}=\cos \theta {\rm
Re}\psi -\sin \theta {\rm Im}\psi 
\]

\begin{equation}
{\rm Im}\psi \longrightarrow {\rm Im}\psi ^{^{\prime }}=\sin \theta {\rm
Re}\psi +\cos \theta {\rm Im}\psi 
\end{equation}

There are three interesting properties in the wave function $\psi $:

(a) The real part ${\rm Re}\psi $ and imaginary part ${\rm Im}\psi $ are
all real numbers, but both of them are nonobservable.

(b) The distinction between ${\rm Re}\psi $ and ${\rm Im}\psi $ is
relative but necessary. One side regards the existence of other side as the
premise of the existence of itself. In other words, they are indivisible,
any one side can not exist alone. A young man Jia Baoyu ( in the classic
Chinese novel ``Dream of the red mansion'' by Cao Xueqin and Gao e around
1760 ) caught an idea one day. He wrote as follows:

You won't be you if I wasn't born.

One can't understand her from her alone.

(c) ${\rm Re}\psi $ and ${\rm Im}\psi $ can transform into each other as
shown in Eq.(2) while keeping $\left| \psi \right| ^2\longrightarrow \left|
\psi ^{^{\prime }}\right| ^2=\left| \psi \right| ^2$ invariant.

We should not overlook these three properties, which go beyond the
``atomism'' familiar to western science community. This is why many
physicists feel QM being too abstract. Actually, this kind of concept
already existed in Chinese philosophy. If we set ${\rm Re}\psi $ and ${\rm %
Im}\psi $ in correspondence with ``yin''( feminine or negative ) and
``yang''( masculine or positive ), then the complex expression of wave
function, its phase transformation and Eq.(2) are nothing but the
mathematical description of the theory of ``Yuan-qi''( the primary gas, see
Ref[2] ). They also correspond precisely to the principles of
``contradiction theory'' in philosophy. The opposition of ``yin'' and
``yang'' comprises the ``repulsiveness'' of contradiction while their mutual
dependence and transformation property comprise the ``identity'' of
contradiction.

\section{Subtlety in the use of $i=\protect\sqrt{-1}$}

The introduction of $i=\sqrt{-1}$ with the property $i^2=-1$ reflects
exactly the law of opposition and mutual transformation of two sides in
contradiction. The use of mathematical language in physics is the most
precise one, it is nearly no substitute. Suppose there is no $i$ in QM, the
situation would become very serious. People would regard the wave function
as some quantity describing the direct observable. That would be totally
wrong. In fact, the development of physics to the stage of QM has been
penetrating into the essence of matter. The essence is nothing but
contradictions, which are invisible and untouchable. The introduction of $i$
in QM just prevents people from creating the extravagant hopes which regard
the wave function as direct observable. On the other hand, it undertakes the
mission to search for the essence of matter successfully. Hence, the quantum
theory is actually a theory of contradiction. Then the question
arises:``What contradiction does the wave function in QM represent?''

Notice that the Sch\"{o}dinger equation is always a linear and homogeneous
equation for $\psi (x,t)$ in various external field $V(x)$ ( in one
dimensional space ). This comprises an acute contrast with the classical
mechanics.

Assume that $V(x)=a_0+a_1x+a_2x^2+a_3x^3+......$ . In classical mechanics, $%
x $ denotes the position coordinate of particle. Then the Newtonian equation 
$F=ma$ with the external force $F=-\frac{dV}{dx}$ and acceleration $a=\frac{%
d^2x}{dt^2}$ leads to equation

\begin{equation}
m\frac{d^2x}{dt^2}+a_1+2a_2x+3a_3x^2+...=0
\end{equation}
$\newline
$If $a_1\neq 0,$ $a_2=a_3=...=0$, Eq.(3) is linear but inhomogeneous. If $%
a_1=0,$ $a_2\neq 0,$ $a_3=a_4=...=0,$ Eq.(3) is linear and homogeneous. If
in general case, $a_3\neq 0,...$ , Eq.(3) is nonlinear.

The reason of appearance of nonlinear equation in classical mechanics lies
in the fact that being the coordinate of particle, $x$ forms directly the
passive (dependent) variable of an ordinary differential equation with one
active (independent)variable $t$ ( time ) only. Things are quite different
in QM, where the $x$ in wave function $\psi (x,t)$ is not the coordinate of
particle, but a flowing coordinate describing the position in space and
forms the active variable together with $t$ . The Schr\"{o}dinger equation
with $\psi $ as its passive variable is always a linear and homogeneous
partial differential equation.

The above fact enlightens us to interpret the wave function $\psi (x,t)$ in
QM as the abstract representation of contradiction of the particle with its
environment and $V(x)$ as the potential energy describing the strength of
interaction between them[3]. Then the linear and homogeneous property of
equation can be explained as the feature of contradiction interaction. The
interaction between particle and its environment is not realized via the
external force pushing forward but an interaction of percolation type------
the external cause works only via the internal cause.

\section{How large is an electron?}

Some reader might think that ``What a contradiction? It is invisible and so
can be talked about at one's disposal. I don't believe in it.'' Good, the
problem is just in nowhere but the extravagant hope to see the essence
directly whereas the latter, i.e., the contradiction is just invisible
before its excitation and transmutation. What we see are phenomena, which
can not be identified as the essence. The latter is always hiding inside and
don't exhibit outside. It only displays itself clearer and clearer with the
increase of excitation energy. Nonetheless, theoretical analysis is needed
for handling the essence.

After an electron is captured by a proton, a Hydrogen atom is formed. An
amount of energy $13.6eV$ , called as binding energy, is delivered.. On the
contrary, if an amount of energy larger than $B$ is injected into the
Hydrogen atom, it will be ionized. According to the theory of Special
Relativity (SR), a free electron has a ``rest energy'' $E_0=m_ec^2=0.511MeV$.

Attention must be paid to the difference between the binding state of an
electron in a Hydrogen atom and its free state. However, because $B/E_0$ is
far less than $1$ , people often neglect the difference between these two
kinds of electron. An electron in free space is described by plane wave
function whereas its wave function of $1s$ state binding in Hydrogen atom
reads $\psi (r)=e^{-r/a}/(\pi a^3)^{1/2}$ with Bohr radius $a=0.0529nm$ and $%
r$ being the distance measured from the nucleus ( proton ).

People often deemed the electron as a point-like particle. If with this
view, it would be very difficult to get rid of the concept that the electron
is in some orbital motion inside the Hydrogen atom, let alone to explain why
it will not emit the electromagnetic wave. After the establishment of QM,
people cognized that the stationary state as shown by the $1s$ wave function
corresponds to some ``de Broglie stationary wave'', which does not emit the
electromagnetic wave and will never fall into the nucleus..

According to the uncertainty relation in QM, the smaller the size of
electron wave packet is compressed, the higher its kinetic energy will be,
i.e., the stronger its tendency against the compression will be. If the
charge number of nucleus increases from $1$ to $Z$, then $a\longrightarrow
a/Z$ in the $1s$ wave function of this Hydrogen-like atom implies the
spreading radius of electron shrinking to a factor of $Z$ , while the
binding energy enhances to a factor of $Z^2$ .

I wish to raise a bold conjecture that an electron is something arbitrary in
size. When it is bound in $1s$ state of Hydrogen-like atom, an electron
shows a spherical distribution with radius $a/Z$ . When it is under the
collision of other high energy particle, an electron behaves itself like a
point particle with radius less than, say, $10^{-16}cm$ .

\section{Two basic points of view in epistemology}

The above claim is based on the following two points of view.

The existence state of an object is depending on its environment. Certain (
many body ) environment determines its property, e.g., its mass,
stability ( life ), etc. The prominent merit of physics in 20th century
lies in the fact that it enables us begin to analyse such kind of problem by
means of the theory like QM.

When we wish to cognize the state of an object, we have to resort to some
measurement. At that time the state of object changes abruptly. Hence the
state of object exhibiting in the measurement is strongly depending on the
style of measurement, i.e., on the device of changing method and its
strength, etc. Though the study of measurement theory in physics is far from
enough, but one thing is beyond any doubt that there is intervention of
apparatus and subject into the measurement. No cognition can be talked about
without subject. An information is created by subject and object in common.
No any data exists before the change in the state of an object occurs in the
measurement.

Therefore, in our point of view, it is groundless to identify the tiny
radius of electron shown in the high energy collision as its size when it is
in low energy state. On the other hand, we should not look at the $x$ or $p$
in the wave function (e.g., the plane wave function or that in $1s$ state )
as the position or momentum of electron before the measurement.

In nonrelativistic QM, a wave packet of freely moving particle will spread
unceasingly, a phenomenon difficult to understand even in the Born
statistical interpretation. In our point of view, the wave function $\psi
(x,t)$ is not the probability amplitude in distribution of position $x$ of
particle but the distribution of ``contradiction field'' under the
interaction between the particle and its environment. One supposes that at $%
t=0$ the wave function $\psi (x,0)$ has a distribution as a wave packet, and
assumes again that $V(x)=0$ i.e., there is no interaction between particle
and environment. Then being the ``contradiction field'', the wave packet has
no way but spreads and approaches to zero. It is totally reasonable, no any
difficulty exists in understanding.

\section{What we learn from the EPR paradox?}

Einstein had made prominent contribution to the establishment of quantum
theory, but he had doubts about QM especially the uncertainty relation. He
doubted first of the consistency of QM, then of its completeness. In 1935 ,
Einstein, Podolsky and Rosen wrote a paper to raise the question whether the
description of QM on the ''physical reality'' is complete. They tried to
explore the possible incompleteness of QM via ideal experiments. (see e.g.,
Ref [4]).

One of the so-called EPR experiments was performed by C.S.Wu et al in 1975.
A positronium, i.e., the binding state of $e^{+}$ and $e^{-}$ will
annihilate into two photons (see Ref [5]):

\begin{equation}
e^{+}+e^{-}\rightarrow \gamma _1+\gamma _2
\end{equation}

The positronium begins at a rest state with spin zero. According to the laws
of conservation of momentum and angular momentum, two photons emitted along
two opposite directions must be either both in right polarization state $%
\mid R_1R_2\rangle $ or both in left polarization state $\mid L_1L_2\rangle .
$ The QM reveals that the final state after decay is a linear superposition
of the above two states:

\qquad 
\begin{equation}
\mid F\rangle =\mid R_1R_2\rangle -\mid L_1L_2\rangle
\end{equation}

There are two observers, say, Bob and Alice . Bob (Alice) performs the
measurement at the west (east) side. Both of them use the linear polarizer
to detect the light. A photon with circular polarization, either left or
right, will have 50\% of probability in either $x$ polarization or $y$
polarization.. According to the calculation of QM, the probability amplitude
that Bob has measured the $\gamma _1$ in $x$ polarization state $\mid
x_1\rangle $ while Alice measured the $\gamma _2$ in $y$ polarization state $%
\mid y_2\rangle $ reads

\qquad 
\begin{equation}
\langle x_1y_2\mid F\rangle =i
\end{equation}

which is understandable. However, the QM predicts that the probability
amplitude for both measurements being in $x$ (or $y)$ polarization equals
zero:

\qquad 
\begin{equation}
\langle x_1x_2\mid F\rangle =\langle y_1y_2\mid F\rangle =0
\end{equation}

From EPR's point of view, the above result seems a violation in the belief
of ''physical reality'' and so is difficult to be accepted. Consider the
original appointment was made to measure the process (6). After $\gamma _1$
and $\gamma _2$ are separated to a long distance, Bob suddenly changes his
mind and measures the $y$ polarization of $\gamma _1,\mid y_1\rangle .$ It
seems that Bob's change should not influence the outcome of Alice's
measurement, i.e., the zero amplitude shown in Eq.(7) can hardly be
understood. However, the prediction of QM does show that in this case Alice
can not find the $\gamma _2$ in $y$ polarization state. Bob's act does
influence Alice's result far away. In the past 60 years, especially after
the study of Bell's Inequality, the accuracy of EPR experiments has been
improved year after year with the distance between Bob and Alice increase
even to 10 Kilometers [6]. The experiments have been verifying the
correctness of prediction of QM and exclude the existence of so-called
''local hidden variable''.

In our point of view, the study on EPR paradox in 60 years has been telling
us one important thing that the quantities appear in the calculation of QM
like the position, the momentum, the angular momentum or spin
orientation,etc., are all not the direct observables. To be precise, they do
not really exist before the measurement. For example, in the entangled
two-photon state, Eq.(5) , before the quantum coherence is destroyed by the
measurement, no one of these two photon is in left polarization or in right
polarization. Similarly, to talk about $x$ polarization or $y$ polarization
is meaningless before the measurement. The reason why we felt strange about
EPR paradox is nothing but that we might look at the ''description'' in QM
too seriously.

\section{The quantum state, wave function, operator and the uncertainty
relation}

According to the rigorous notation of Dirac, a state of microscopic particle
is denoted by an abstract state vector $\mid \Psi \rangle $, which even does
not contain the time $t$ in the Heisenberg picture, let alone $x$ or $p$.
Then the problem of the choice of representation arises. One may choose the
basic vector of position $\mid x,t\rangle $ to get the wave function in
configuration space , i.e., in $x$ representation as

\begin{equation}
\Psi (x,t)=\langle x,t\mid \Psi \rangle
\end{equation}

Alternatively, one may choose the basic vector of momentum $\mid p,t\rangle $
to get the wave function in momentum space , i.e., in $p$ representation as

\begin{equation}
\Phi (p,t)=\langle p,t\mid \Psi \rangle
\end{equation}

Neither $x$ nor $p$ is more fundamental than the other one. Actually, they
are not really existing in the original state $\mid \Psi \rangle $ as the
position or momentum of particle. At most, we could say that $x$ or $p$
would be the position or momentum of the particle found by us if the
measurement on the state $\mid \Psi \rangle $ were really made.

In our point of view, the choice of basic vector $\mid x,t\rangle $ or $\mid
p,t\rangle $ for representation corresponds to the choice of apparatus for $%
x $ or $p$ measurement we are going to use. However, before the measurement
is really done, it is just a fictitious one. Hence the corresponding wave
function, i.e., the probability amplitude contains the $i=\sqrt{-1},$showing
its unobservableness. This is why we explain the wave function as a
''contradiction field'' to stress its unobservableness before the
transmutation of contradictions.

Once the measurement is made, it is a changing (operating) process. The
transmutation of contradiction occurs immediately. For instance, the
variable $p$ in the plane wave function for free electron $\Psi \sim \exp
\{\frac i\hbar (p\cdot x-E\cdot t)\}$ is not a directly observable momentum
of electron, but showing a potential possibility that under an operation of
infinitesimal ''space translation'' (i.e., under an infinitesimal change in $%
x$ ), an observable $p$ would emerge:

\begin{equation}
-i\hbar \lim_{\Delta x\rightarrow o} \frac{\Psi (x+\Delta
x,t)-\Psi (x,t)}{\Delta x}=-i\hbar \frac \partial {\partial x}\Psi =p\Psi
(x,t)
\end{equation}

Now we understand why an observable (like $p$ ) in classical physics evolves
into an operator in QM ( like $\stackrel{\wedge }{p}=-i\hbar \frac \partial
{\partial x}$). In our point of view, it is just a reflection of principle
in epistemology that ''the measurement is nothing but an operation (
changing ) on the object''.

this principle is by no means beginning from QM. We already had it in
classical physics. But people often overlooked it. We had discussed as an
example the measurement of specific heat under constant volume $c_v=\frac{%
\partial U}{\partial T}\mid _v$ or that under constant pressure $c_p=T\frac{%
\partial S}{\partial T}\mid _p,$ see Ref [6].

Notice that, however, the operation of space translation during the process
for measuring the momentum $p$ is bound to disturb the measurement of
position $x$ . The nonconformity between two kinds of operation leads to the
noncommutativity between these two operators and also to the uncertainty
relation. Even in classical physics, we would meet the similar situation
when measuring the $c_v$ and $c_p$ at the same time.

In summary,we have tried to answer the question raised at the beginning of
this article:why the two aspects of subtlety------the noncommutativity and
the existence of probability amplitude------are correlated intimately. The
reason lies in the fact that any observable is nothing but a consequence of
transmutation in contradiction induced in a changing process. The
'contradiction'' is characterized by $i=\sqrt{-1}$ and its transmutation
characterized by the Planck constant $\hbar .$

\section{A particle is an excitation state of fundamental contradiction in
nature}

At the level of QM, the contradiction is relatively weak and the change of
particle is small. For instance, one often ignores the distinction between
the free electron and the electron inside a Hydrogen atom. Once when the
energy exceeds $E_0$ and so the electron itself could be created or
annihilated, i.e., at the level of quantum field theory, we are facing
directly the excitation and transmutation of fundamental contradictions in
nature.

As an example, the collision between electron ($e^{-}$ ) and positron ($e^{+}
$) would create a $J/\psi $, a meson. The rest mass of $J/\psi $ ($%
3.097GeV/c^2$) is 6000 times as much as that of an electron ($0.511MeV/c^2$).
How can one say that $J/\psi $ is hiding originally inside the $e^{+}$ or $%
e^{-}$? We prefer to say that $J/\psi $ is excited from the vacuum. Usually,
it is considered as a binding state of a pair of quark-antiquark, $J/\psi
=(c-\overline{c})$. However, $c$ or $\overline{c}$ has never been detected
in free state. If the energy is raised into $2\times 1.868GeV$, the ''bond''
between $c$ and $\overline{c}$ is broken while a new quark pair, $d$ and $%
\overline{d}$, is created from the vacuum to form two separated $D$ mesons, $%
D^{+}=(c-\overline{d}$ $)$, $D^{-}=(\overline{c}-d).$ The isolated quark has
never been seen and is called as the confinement of quark..

In the beginning of 20th century, people were bothered by the puzzle ''why
an electron will never fall into the nucleus?'' Now people feel strange
again by the puzzle ''why the quark can not escape from the hadron?'' It
seems that a radical change in the concept of matter structure is
inevitable. We can not prevent from querying that as people had
underestimated the repulsiveness of contradiction again and again, if we may
still underestimate the identity of it today?

In comparing with the discussion on the real and imaginary parts of wave
function, we guess that the quark may be some representation of fundamental
contradiction. While quark $c$ ($\overline{c})$ has one character, $d$ ($%
\overline{d}$) has another one, both of them are different from that of $%
e^{-}(e^{+}).$ According to our present cognition, the difference between
the wave functions of $e^{-}$ and $e^{+}$ is only ascribed to the opposite
sign in its phase.

\begin{eqnarray}
\Psi _{e^{-}} &\sim &\exp \{\frac i\hbar (\overrightarrow{p}\cdot 
\overrightarrow{x}-E\cdot t)\}  \nonumber \\
\Psi _{e^{+}} &\sim &\exp \{-\frac i\hbar (\overrightarrow{p}\cdot 
\overrightarrow{x}-E\cdot t)\}  
\end{eqnarray}

($E>0$ in both case). Starting from this point of view, we are able to
derive the main results of SR. Also we may understand why the wave packet of
an electron in Dirac theory will not spread when $v\rightarrow c.$ Rather,
it undergoes a boosting effect accompanying the Lorentz contraction. This is
due to the entanglement of two kinds of contradiction field, that of
particle and of antiparticle, inside an electron.

\section{Will the ''Ether'' theory revive?}

The development of QM enables us try to answer the question raised by
Einstein :''what is the physical reality?'' It seems to us that the matter
needs to be defined at two levels. When an object is independent of the
consciousness of mankind and before the measurement is made, it could be
called as ''thing in itself''. In QM it is denoted by a quantum state $\mid
\Psi \rangle $ separated approximately from its environment and contains no
information. Then when it appears as various phenomena, reflecting a series
of experimental data in our measurement, it is turned into ''thing for
us''.The wave functions in QM just connects the quantum state to phenomena
via fictitious measurements. As wheeler said:''No phenomenon is phenomenon
until it is an observed phenomenon .''

In Fig 1 (see the original paper in {\it Kexue} (Science) Vol. 50 No. 2,
38, 1998 or the similiar one in Internet hep-th/9508069),
a particle is shown as certain excitation of the vacuum, which may
also be called as ''Yuan-qi'' (the primary gas, [2] ) or ''Ether''.The lower
part of Fig 1 is not directly observable. But after the excitation of
particle with momentum ($\overrightarrow{p}$) and energy ($\ E$) as its
existence form, we may also say that the space ($\overrightarrow{x})$ and
time ($t$) are the existence form of the vacuum. An universal constant, the
speed of light $c$, exhibits itself as the horizontal link. On the other
hand, the vertical excitation is displayed by the quantum operator
relationship:

\begin{eqnarray}
\overrightarrow{p} &=&-i\hbar \nabla,    ~~~~~~~~~~~~\overrightarrow{p_c}%
=i\hbar \nabla  \nonumber \\
E &=&i\hbar \frac \partial {\partial t}, ~~~~~~~~~~~~E_c=-i\hbar \frac
\partial {\partial t}  
\end{eqnarray}

where the Planck constant $\hbar $ exhibits as the vertical link. The left
two relations in Eq.(12) are well known in QM whereas the right two with
subscript $c$, which belong to the antiparticle (see Eq.(11)), are provided
by SR in our point of view. Eq.(12) as a whole
are just the DNA in quantum field theory and particle physics inherited from
QM and SR each with 50\% [8].

It seems to us that the revival of a new ''Ether'' theory based on
''Yuan-qi'' is inevitable. It is the time to fuse two kinds of wisdom of
both eastern and western. Sooner or later, we will be convinced by the
saying in Chinese philosophy:''Oneness of heaven and man''.



\begin{thebibliography}{99}

\bibitem{1} C.N.Yang, Schr\"{o}dinger Centenary Celebration of a Polymath, ed. by
C.W.Kilmister, Cambridge Univ. Press. 1987

\bibitem{2} Z-x He, Philosophical thinking on the quantum composite field theory.
Beijing Normal Univ. Press 1997, 275

\bibitem{3} G-j Ni and H-f Li, Modern physics, Shanghai Science \& Technology Press,
1979

\bibitem{4} R-K Su, Quantum Mechanics, Fudan Univ. Press 1997

\bibitem{5} R.P.Feynman, R.B.Leighton,M.Sands, The Feynman Lectures on Physics
3------ Quantum Mechanics, Addison-Wesley Publishing Co. Inc., 1965

\bibitem{6} W.Tittel et al., Preprint, Internet, quant-ph/9707042, To appear in Phys.
Rev. A

\bibitem{7} G-j Ni, Measurement and epistemological problems in contemporary physics,
Philosophy Research, 1978, April, 63.

\bibitem{8} G-j Ni, Kexue (Science) $\underline{50}$, No.1,29 (1998), Internet
quant.Phys/9803034

\end{thebibliography}
\end{document}